\documentclass{emulateapj}

\newcommand {\kms}{\,\rm km \, s^{-1}}

\begin{document}

\title{Modeling kicks from the merger of generic black-hole binaries}
\author{John G. Baker, William~D.~Boggs \altaffilmark{1}, Joan Centrella, Bernard~J.~Kelly,
Sean T. McWilliams \altaffilmark{1}, M.~Coleman Miller \altaffilmark{2}, James~R. van~Meter}
\affil{Laboratory for Gravitational Astrophysics, 
NASA Goddard Space Flight
Center, Greenbelt, Maryland 20771}
\altaffiltext{1}{University of Maryland, Department of Physics, College
  Park, Maryland 20742-4111}
\altaffiltext{2}{University of Maryland, Department of Astronomy,
College Park, Maryland 20742-2421}

\begin{abstract}
  Recent numerical relativistic results demonstrate that the merger of
  comparable-mass spinning black holes has a maximum ``recoil kick''
  of up to $\sim 4000 \kms$.  However the scaling of these recoil
  velocities with mass ratio is poorly understood. We present new runs
  showing that the maximum possible kick perpendicular to the orbital
  plane does not scale as $\sim\eta^2$ (where $\eta$ is the symmetric
  mass ratio), as previously proposed, but is more consistent with
  $\sim\eta^3$, at least for systems with low orbital precession.  We
  discuss the effect of this dependence on galactic ejection scenarios
  and retention of intermediate-mass black holes in globular clusters.
\end{abstract}

\keywords{black hole physics -- galaxies: nuclei -- gravitational waves --- relativity }

\section{Introduction}

Recently, numerical exploration of the radiative recoil ``kick'' of
merging black holes has progressed considerably.  In particular,
efforts in this regard have led to suggested phenomenological formulae
for the kick, largely based on post-Newtonian (PN) predictions such as
that given by \cite{Kidder95a}, which have proved surprisingly
successful.  For example, \cite{Gonzalez:2006md} found that in cases
of unequal masses ($q \equiv m_1/m_2 < 1$) and no spin, a simple
modification of the PN formula originally found by
\cite{Fitchett_1983} fits the numerical data quite well.  For cases of
spins perpendicular to the orbital plane (i.e.\ parallel with the
orbital angular momentum), a formula proposed by \cite{Baker:2007gi}
is also consistent with numerical data.  This formula is loosely based
on PN calculations, with spins perpendicular to the orbital plane
producing kicks in the orbital plane.  For spins with components in
the orbital plane, \cite{Campanelli:2007ew} have proposed a formula,
again derived from PN calculations, that agrees well with numerical
results for equal masses.

This last type of kick, which is perpendicular to the orbital plane,
is of particular interest because its computed magnitude can be very
large (up to thousands of kilometers per second). In the current
literature (specifically \citealt{Campanelli:2007ew}), the mass-ratio
dependence is drawn from the leading-order PN approximation.  It is
unclear whether this approximation is sufficient to predict the
strong-field dynamics that presumably determines the kick.  Indeed,
hints of a deviation from this form are evident for mass ratio $q =
1/2$ in the runs of \cite{Lousto:2007db}.  Therefore, although the
angular dependence of the proposed formula is consistent with symmetry
arguments, which are independent of the strong-field dynamics
\citep{Boyle:2007sz,Boyle:2007ru}, the mass ratio dependence of this
formula is currently not well justified.

Characterization of the dominant kick for unequal masses is especially
important because, although the largest possible kicks would eject the
remnant from any galaxy, for astrophysical applications the
distribution of kick speeds matters most. For example,
\cite{Bonning:2007vt} find no evidence for quasars ejected from their
hosts (although \citealt{Komossa:2008qd} may have seen a $2650 \kms$
kick). If quasar activity is commonly induced by major galaxy mergers
that lead to coalescence of supermassive black holes, the implications
of this therefore depend in part on how frequently one expects a
merger to allow ejection.  Even for very large recoils, more than half
of galaxies would still retain black holes at their cores
(\citealt{Schnittman:2007nb}).  However, the kick speed distribution
has a major impact on the hierarchical growth of massive black holes
at redshifts $z>5$ (e.g., \citealt{Volonteri:2007et}).

Here we investigate how the out-of-plane kick depends on the mass
ratio, and find that, for mass ratios in the range $q=1$ to $q=1/3$
and spins $S_i\leq 0.2m_i^2$, the kick drops off more rapidly with
decreasing mass ratio than proposed by \cite{Campanelli:2007ew}.
Specifically, we find that a large body of numerical data on kicks are
well represented by
\begin{eqnarray}
\vec{V}_{\rm recoil} &=& v_m \, \hat{e}_1 + v_{\perp} (\cos\xi \, \hat{e}_1 + \sin\xi \, \hat{e}_2) + v_{\parallel} \, \hat{e}_2, \label{eq:v_total}\\
      v_m     &=& A \eta^2 \sqrt{1 - 4 \eta} (1 + B \eta), \label{eq:v_mass}\\
v_{\perp}     &=& H \frac{\eta^2}{(1+q)} \left( \alpha_2^{\parallel} - q \alpha_1^{\parallel} \right), \label{eq:v_perp}\\
v_{\parallel} &=& \frac{K \eta^3}{(1+q)}\left[q\alpha_1^{\perp} \cos(\phi_1-\Phi_1) - \alpha_2^{\perp} \cos(\phi_2-\Phi_2)\right]\,, \label{eq:v_parallel}
\end{eqnarray}
where $v_m$ is the mass asymmetry contribution, $v_{\perp}$ and
$v_{\parallel}$ are the spin contributions that yield kicks
perpendicular and parallel to the orbital angular momentum.
$\eta\equiv q/(1+q)^2$ is the symmetric mass
ratio. $\alpha_i^{\parallel}$ is the projection of the dimensionless
spin vector $\vec{\alpha}_i=\vec{S}_i/m_i^2$ of black hole $i$ along
the orbital angular momentum, while $\alpha_i^{\perp}$ and $\phi_i$
are the magnitude and angle with respect to some reference angle in
the orbital plane of its projection, $\vec{\alpha}_i^{\perp}$, into
the orbital plane.  $\Phi_1$ and $\Phi_2$ are constants for a given
mass ratio.  Here, $A = 1.35 \times 10^4 \kms$, $B = -1.48$, $H = 7540
\pm 160 \kms$, $\xi = 215^\circ \pm 5^\circ$, and $K = 2.4 \pm 0.4
\times 10^5 \kms$.  This formula, similar in form to that of
\cite{Campanelli:2007ew}, synthesizes results from
\cite{Gonzalez:2006md} for (\ref{eq:v_mass}) and from
\cite{Baker:2007gi} for (\ref{eq:v_total}) and (\ref{eq:v_perp})
\footnote{Note that in \cite{Baker:2007gi}, we used a simpler form for
the zero-spin contribution, equivalent to (\ref{eq:v_mass}) with
$B=0$.}. For $\xi$ and $H$ we have fit available numerical data from
\cite{Herrmann:2007ac,Koppitz:2007ev,Baker:2007gi}.  The qualitatively
new part, the factor of $\eta^3$ in (\ref{eq:v_parallel}), replaces
the factor of $\eta^2$ originally proposed by
\cite{Campanelli:2007ew}, and is motivated by new numerical evolutions
presented here.

\section{Initial Data and Methodology}
\label{sec:method}

\begin{figure}
\begin{center} 
\plotone{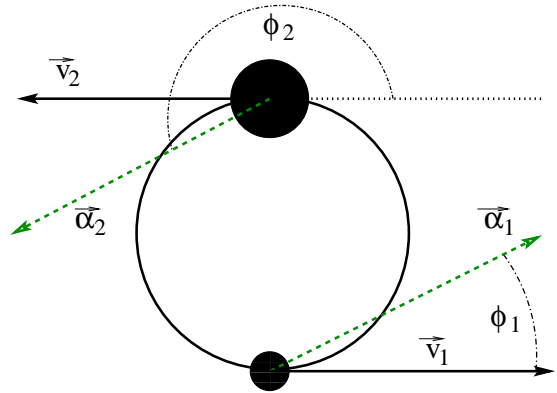}
\caption{Configuration of black holes for all new simulations. The two
  holes' spins $\vec{\alpha}_{(1,2)}$ lie initially in the orbital
  plane, at angles $\phi_1$ and $\phi_2$ to $\vec{v}_1$, the smaller
  hole's initial velocity.}
\label{fig:BHconfig}
\end{center}
\end{figure}

We simulated the inspiral and merger of a range of spinning black-hole
binaries, with mass ratios in the range $1/1.1 \geq q \geq 1/3$. The
initial configuration of momenta and spins is illustrated in
Fig.~\ref{fig:BHconfig}. The parameters used in the numerical
evolutions are presented in the first three columns of Table
\ref{tab:initial}. For these evolutions, the smaller hole ($m_1$)
has a dimensionless spin $|\vec{\alpha}_1| = 0.2$, while the larger
hole's spin is $|\vec{\alpha}_2| = q^2 |\vec{\alpha}_1|$. Both spins
initially lie in the orbital plane, at angles $\phi_1$ and
$\phi_2$ to the initial velocity of hole 1 (see Fig.~\ref{fig:BHconfig}).

\begin{deluxetable}{lrr|rr}
\tablecolumns{5}
\tablewidth{0pt}
\tablecaption{Initial Configuration and Final Kick for Each Simulation.
$\phi_{1(2)}$ is the angle made by the spin vector of hole 1(2) with
the velocity vector of hole 1, as shown in
Fig. \ref{fig:BHconfig}. Numerical results for the kick components
$v_m$ (where available) and $v_{\parallel}$ are shown.  Kicks for
equivalent spinless runs are in parentheses.\label{tab:initial}}
\tablehead{\colhead{$q$} & \colhead{$\phi_1 (\,^o)$} & \colhead{$\phi_2 (\,^o)$} & \colhead{$v_m$} ($\kms$) & \colhead{$v_{\parallel}$ ($\kms$)}}
\startdata
 1/1.1   &   0 & 180 & 24 & -542 \\
         & 315 & 135 & 24 & -657 \\
         & 270 &  90 & 25 & -384 \\
&&&&\\   
 1/1.3   &   0 & 180 & 67 & -386 \\
         & 315 & 135 & 67 & -525 \\
         & 270 &  90 & 69 & -348 \\
&&&&\\   
 1/1.5   &  60 & 240 & 92 (94) & -381 \\
         &   0 & 180 & 95 (94) & -135 \\
         & 315 & 135 & 91 (94) &  168 \\
         & 270 &  90 & 90 (94) &  364 \\
&&&&\\   
1/2      &   0 & 180 & 137 (140) &  -37 \\
         & 315 & 135 & 136 (140) &  111 \\
         & 270 &  90 & 136 (140) &  193 \\
         & 315 &  90 & \nodata &   75 \\
         &   0 &  90 & \nodata &  -55 \\
&&&&\\   
1/3      &   0 & 180 & 166  &	49 \\
         & 315 & 135 & 166  &	48 \\
         & 270 &  90 & 163  &	17 \\
         &   0 &   0 & 162  &  114 \\
\enddata
\end{deluxetable}

To perform our simulations, we employed the \textsc{Hahndol} evolution
code, as described by \cite{Baker:2007gi,BakerTBA}.  We chose initial
coordinate separations of $7.0M$ for the $q\geq 1/2$ cases and $8.0M$
for the $q=1/3$ cases (where $M$ is the total mass of the system) to
yield between one and four orbits prior to merger; informed by PN
theory \citep{Damour:2000we}, we chose the corresponding momenta to
minimize initial eccentricity.  The finest grid spacing in all the
runs presented here was $h_f=3M/160$. We also performed a single
high-resolution simulation of $h_f=M/64$ for the $q=1/2$ case and
found the kicks and all other relevant quantities agreed with the
corresponding $h_f=3M/160$ simulation to within $\sim 1$\%;
additionally we performed a set of lower-resolution $h_f=3M/128$
simulations in the $q=1/3$ case, demonstrating consistency of the
amplitude of $v_{\parallel}$ with the $h_f=3M/160$ simulations to
within 6\%.

\section{Results and Discussion}
\label{sec:results}

The recoil kicks resulting from the new simulations are in the
rightmost columns of Table~\ref{tab:initial}. Note that the in-plane
kick agrees well with what would have been expected given no spin
($v_m$). This indicates negligible orbital precession, which we also
verified from the trajectories of the black hole centers.

To conceive of plausible candidates for the mass scaling of
$v_{\parallel}$, we begin with the spin expansion and symmetry
arguments of \cite{Boyle:2007sz,Boyle:2007ru}.  For the spin
configurations considered here,
\begin{equation}
v_{\parallel}=D(q)\alpha_1^{\perp}\cos\left(\phi_1-\Phi(q)\right)-D(1/q)\alpha_2^{\perp}\cos\left(\phi_2-\Phi(1/q)\right),
\end{equation}
where $D$ and $\Phi$ are some functions of mass ratio $q$, and we note
that $\Phi$ must also depend on the initial separation. Further
restricting ourselves to forms relatable to the factor of
$\vec{S}_1/m_1-\vec{S}_2/m_2$ appearing in PN calculations of the
kick, which informed \cite{Campanelli:2007ew,Lousto:2007db} and has
been numerically well-verified in the equal-mass case, we substitute
$D(q)=qC(\eta)/(1+q)$ to obtain:
\begin{equation}
\label{eq:generalized_kick}
v_{\parallel} = \frac{C(\eta)}{(1+q)}\left[q\alpha_1^{\perp} \cos(\phi_1-\Phi_1) - \alpha_2^{\perp} \cos(\phi_2-\Phi_2)\right],
\end{equation}
where $\Phi_1\equiv\Phi(q)$ and $\Phi_2\equiv\Phi(1/q)$.
Eq.~\ref{eq:v_parallel} arises from the choice $C(\eta) = K \eta^3$,
with $K = 2.4 \times 10^5$.

There are several other possibilities for the form of $C(\eta)$ in the
literature. \cite{Campanelli:2007ew} assume $C(\eta) = 6.0\times10^4
\eta^2$. In this case, $(\phi_1 - \Phi_1)$ and $(\phi_2 - \Phi_2)$ are
related to $(\Theta - \Theta_0)$ from \cite{Campanelli:2007ew}, as
well as the angle between the spin vectors implicit in
$|\vec{\alpha}_2^{\perp}-q\vec{\alpha}_1^{\perp}|$.

Another possibility arises from the known relation between
$v_{\parallel}$ and the difference between the energy radiated in the
$(l,m)=(2,2)$ and $(2,-2)$ harmonics of the radiation
\citep{Schnittman:2007ij,Brugmann:2007zj}.  With no spin, these
quantities are equal.  With spin, we expect that $v_{\parallel}\sim
\dot{E}_{22(peak)}F$, where $\dot{E}_{22(peak)}$ is the peak power
radiated in the $(2,2)$ harmonic, and $F$ represents the
spin-dependent asymmetry between $\dot{E}_{22(peak)}$ and
$\dot{E}_{2-2(peak)}$, i.e.  $F\sim
1-\dot{E}_{2-2(peak)}/\dot{E}_{22(peak)}$.  For black holes with no
spin, we have found that $\dot{E}_{22(peak)}= a_2\eta^2+a_4\eta^4$,
where $a_2=0.0044$ and $a_4=0.0543$, gives a good fit to the numerical
data \citep{Baker:2008mj}.  We do not expect spins orthogonal to the
orbital angular momentum to change the scaling of the radiated energy
significantly.  If we further assume that the asymmetry factor $F$ is
independent of $\eta$, which finds some support in PN analysis since
to leading order $\dot{P}_{\parallel}/\dot{E}$ is independent of
$\eta$, then we hypothesize that
$C(\eta)\propto(a_2\eta^2+a_4\eta^4)$.

\begin{deluxetable}{l|r|r|r}
\tablecolumns{4}
\tablewidth{0pt}
\tablecaption{Maximum percent error resulting from various models of the kick, as distinguished by overall mass-ratio dependence.  See equation (8).
\label{tab:fits}}
\tablehead{\colhead{$q$} & \colhead{ $K\eta^2$ } & \colhead{$K(a_2\eta^2+a_4\eta^4)$} & \colhead{$K\eta^3$}}
\startdata
         1/1.1 & \quad \qquad \qquad 0.22 & 0.23 & \quad \qquad \qquad 0.20 \\
&&&\\   
         1/1.3 &  0.75 & 0.80 & 0.78 \\
&&&\\   
         1/1.5 &  1.26 & 1.31 & 1.28 \\
&&&\\   
         1/2   & 15.57 & 2.46 & 1.41 \\
&&&\\   
         1/3   & 39.58 & 10.98 & 9.11 \\
\enddata
\end{deluxetable}

In Table~\ref{tab:fits} we summarize the agreement of various kick
formulas with the numerical data.  For each formula, which has the
form of Eq.~(\ref{eq:generalized_kick}), we found the best $\Phi_1$
and $\Phi_2$, per mass ratio, according to a least-squares fit to the
data given in Table~\ref{tab:initial}.  For each mass ratio, the
resulting percent error is given for each model, maximized across
initial angle.  Referring to Eq.~(\ref{eq:generalized_kick}), the
column headings $K\eta^2$, $K(a_2\eta^2+a_4\eta^4)$ and $K\eta^3$ of
Table~\ref{tab:fits} represent choices for $C(\eta)$ that were tested,
where in each case $K$ has been chosen so as to reproduce the value of
the formula of \cite{Campanelli:2007ew} in the equal-mass case.

Now we consider the agreement of our data with the $C(\eta) = K
\eta^2$ scaling of \cite{Campanelli:2007ew} (first column of
Table~\ref{tab:fits}).  The error of the best fit grows significantly
with mass ratio, hence the mass-ratio-dependence of this formula is
inaccurate.  One might suppose that precession of the spins into the
orbital plane could account for this.  However, the $v_m$ column in
Table~\ref{tab:initial} shows that the in-plane kicks are close to
those measured without spins (given in parentheses); hence this does
not explain the discrepancy in $v_\parallel$ from the $\eta^2$
scaling. We have experimented with other values of $K$ to resolve the
discrepancy. For example, the maximum error of the $\eta^2$ model can
be reduced to less than $10\%$ for the $q=1/3$ case, but not without
increasing the maximum error of other mass ratios closer to unity to
significantly greater than $10\%$.

Since original submission of this paper, new data presented by
\cite{Lousto:2008dn} seem to indicate $\eta^2$ scaling, although the
cases analyzed are complicated by considerable orbital precession. For
example, their in-plane kicks are apparently at odds with previous
formulae.  It is possible that strongly precessing orbits require
different fitting formulae, but this has yet to be settled.

The choice $C(\eta)=K(a_2\eta^2+a_4\eta^4)$, motivated above, fits the
data much more successfully (second column of Table~\ref{tab:fits}).
Other scalings can be motivated through post-Newtonian-based analysis
\citep{Schnittman:2007ij}. However, a better empirical model was found to be $C(\eta)=K\eta^3$
(third column of Table~\ref{tab:fits}).  For now we consider this our
best fit, and leave open the interesting question of how to accurately
relate this prefactor directly to $\dot{E}_{22}$.

Our results affect the distribution of kick speeds, given various
assumptions about the spin parameters, spin orientations, and mass
ratios involved in coalescences.  This has particular application to
the retention of the products of mergers of massive black holes in the
current universe (e.g., \citealt{Bonning:2007vt}) and electromagnetic
signatures of kicks (e.g., \citealt{Shields:2007ca,Lippai:2008fx}), as
well as coalescences in the early structure formation phase of
redshift $z\sim 5-30$
\citep{Merritt:2004xa,Boylan-Kolchin:2004tf,Haiman:2004ve,Madau:2004st,
Yoo:2004ze,Volonteri:2005pn,Libeskind:2005eh,Micic:2005gj,Volonteri:2007et}, 
and for current-day mergers of intermediate-mass
black holes (IMBHs), which might exist in dense stellar clusters 
\citep{Taniguchi:2000mp,Miller:2001ez,Miller:2002pg,Mouri:2002mc,
Mouri:2002mw,Miller:2003sc,Gultekin:2004pm,Gultekin:2005fd,
O'Leary:2005tb,O'Leary:2007qa}.  Note that $q=1$ to $q=1/3$ is in the
range of ratios expected for major mergers of galaxies, and as
\cite{Sesana:2004sp} show, this range is expected to account for most 
massive black hole mergers in the early $z>10$ phase of black hole
assembly.

Our new formula implies an important revision in our understanding of
how easily IMBHs with $M\sim 10^2-10^3~M_\odot$ are retained in
globular clusters.  A rich cluster has an escape speed $v_{\rm
esc}\approx 50\kms$ \citep{Webbink85}. \cite{Gultekin:2005fd} showed
that the Newtonian kicks involved in binary-single interactions are
insufficient to reach this speed if the IMBH is at least $\sim 15-20$
times more massive than the objects with which it interacts.  Using
the \cite{Campanelli:2007ew} formula, however, the maximum kick from
gravitational radiation is $v_{\rm max}=6\times 10^4 \kms \eta^2$,
implying that even IMBHs $30-35$ times more massive than the black
holes with which they merge could get ejected.
\cite{HolleyBockelmann:2007eh}, focusing on cases in which stars lose
little mass through their evolution and thus can leave behind
stellar-mass black holes with masses $>60-100~M_\odot$, use this to
argue that most IMBHs of even $1000~M_\odot$ will be ejected from
globulars.  If instead stellar-mass black holes have masses $\sim
10~M_\odot$, a mass of at least $400~M_\odot$ would still be required
to guarantee retention.

In contrast, our new formula suggests a maximum kick of $v_{\rm
max}=2.4\times 10^5 \kms \eta^3$.  Thus if $\eta<0.06$, $v_{\rm
max}<50 \kms$.  Therefore, an IMBH interacting with $10~M_\odot$ black
holes will stay in a rich globular if its initial mass is
$M>170~M_\odot$, comparable to what is necessary for retention against
Newtonian three-body kicks.

Our results also have implications for whether merged supermassive
black holes stay in their host galaxies.  The figure of merit is the
fraction of kicks that exceed typical escape speeds from galactic
centers (ranging from roughly $500~\kms$ for a small spiral to
$2000~\kms$ for a giant elliptical), given assumptions about the
distribution of spins and orbital orientations.  The calculation of
record for this purpose is that by \cite{Schnittman:2007sn}, who used
a kick formula based on effective one-body analysis and is different
from that of \cite{Campanelli:2007ew}; this formula underestimates the
highest kicks.  Table~\ref{tab:kickcomp} compares the fraction of
kicks above $500~\kms$ and $1000~\kms$ using the
\cite{Schnittman:2007sn} formula (an underestimate), the
\cite{Campanelli:2007ew} formula (an overestimate), and our results.
It is clear that the \cite{Schnittman:2007sn} results were
conservative: the fraction of large kicks is significantly higher than
their estimate for comparable-mass mergers with plausible spins.

\begin{deluxetable*}{lrrrr}
\tabletypesize{\small}
\tablecolumns{4}
\tablewidth{0pt}
\tablecaption{Fraction of kick speeds above a given threshold, compared
with the results of \cite{Schnittman:2007sn} (SB) and \cite{Campanelli:2007ew}
(CLZM). In all cases we assume
an isotropic distribution of spin orientations.\label{tab:kickcomp}}
\tablehead{\colhead{Mass ratio and spin} & \colhead{Speed threshold} & \colhead{SB} & \colhead{CLZM} &\colhead{This work}}
\startdata
$1/10 \leq q \leq 1$, $a_1=a_2=0.9$        & $v>500 \kms$  & 
$0.12 (+0.06, -0.05)$    & 0.364$\pm$0.0048  & 0.2283$\pm$0.0014\\
                                             & $v>1000 \kms$ & $0.027 
(+0.021, -0.014)$ & 0.127$\pm$0.0034  & 0.085$\pm$0.0008 \\
$1/4 \leq q \leq 1$, $a_1=a_2=0.9$         & $v>500 \kms$ & $0.31 
(+0.13, -0.12)$    & 0.699$\pm$0.0045  & 0.618$\pm$0.0014\\
                                             & $v>1000 \kms$ & 
$0.079 (+0.062, -0.042)$ & 0.364$\pm$0.0046  & 0.2547$\pm$0.0013 \\
$1/4\leq q \leq 1$, $0\leq a_1,a_2\leq 1$ & $v>500 \kms$  &   \nodata               & 0.428$\pm$0.0045  & 0.3484$\pm$0.0015\\
                                             & $v>1000 \kms$ &  \nodata               & 0.142$\pm$0.0034  & 0.0974$\pm$0.0009\\
\enddata
\end{deluxetable*}

Barring mechanisms to retain supermassive black holes after major
mergers, one would expect tens of percent of merged galaxies to have
no central black hole, in strong contradiction with observations (see
\citealt{Ferrarese05}).  Low spin magnitudes would lower kicks, but
this is contrary to spin inferences from Fe~K$\alpha$ lines; see
\cite{Iwasawa96,Fabian:2002gj,Reynolds:2002np,Brenneman:2006hw}.
Alignment of spins is another possibility; since pure gravity does
not do this \citep{Schnittman:2004vq,Bogdanovic:2007hp}, external
torques such as those from nuclear gas would be required
\citep{Bogdanovic:2007hp}.

In conclusion, we have performed a systematic study of the mass ratio
dependence of the out-of-plane kicks produced by the merger of
spinning black holes.  Our work shows that the
\cite{Campanelli:2007ew} candidate kick formula overestimates the
out-of-plane kick systematically. However, we find that an additional
factor of $4\eta$ agrees with our numerical results to within 10\%
(and typically $\sim 1$\%) for mass ratios between 1 and 1/3.  This
has considerable implications for black hole retention in early dark
matter halos, galaxies, and globular clusters.

\acknowledgments

The work at Goddard was supported in part by NASA grant
05-BEFS-05-0044 and 06-BEFS06-19. The simulations were carried out
using Project Columbia at the NASA Advanced Supercomputing Division
(Ames Research Center) and at the NASA Center for Computational
Sciences (Goddard Space Flight Center). B.J.K. was supported by the
NASA Postdoctoral Program at the Oak Ridge Associated
Universities. S.T.M. was supported in part by the Leon A. Herreid
Graduate Fellowship. MCM gratefully acknowledges support from the NSF
under grant AST 06-07428.

%\bibliographystyle{../bibtex/aj}
%\bibliography{../bibtex/references}

\end{document}